# Microwave and RF signal processing based on integrated soliton crystal optical microcombs

Xingyuan Xu, Mengxi Tan, Jiayang Wu, Roberto Morandotti, Arnan Mitchell, and David J. Moss

*Abstract*—**Microcombs are powerful tools as sources of multiple wavelength channels for photonic RF signal processing. They offer a compact device footprint, large numbers of wavelengths, and wide Nyquist bands. Here, we review recent progress on microcomb-based photonic RF signal processors, including true time delays, reconfigurable filters, Hilbert transformers, differentiators, and channelizers. The strong potential of optical micro-combs for RF photonics applications in terms of functions and integrability is also discussed.**

## I. INTRODUCTION

PHOTONIC RF techniques have attracted great interest during the past two decades since they offer ultra-high RF bandwidths, low transmission loss and strong immunity to electromagnetic interference. They have found wide applications ranging from radar systems to communications [1-5]. Among the many useful photonic RF techniques, optical frequency combs are one of the most fundamental and powerful tools due to their ability to provide multiple wavelength channels that can greatly increase the capacity of communications systems and allow the broadcast of RF spectra for advanced signal processing functions. However, traditional approaches for frequency comb generation, including those based on discrete laser arrays or electro-optic (EO) modulation, [6-9] all face limitations, such as their bulky size and large cost brought about by laser arrays and RF sources, or a limited free spectral range (FSR) of the comb lines that in turn yields a limited Nyquist zone for the RF system.

Kerr optical frequency combs [10-22], or "microcombs", that originated from the optical parametric oscillation in monolithic micro-ring resonators (MRRs), offer distinct advantages over traditional multi-wavelength sources for RF applications, such as the potential to provide a much higher number of wavelengths, an ultra-large FSR, as well as greatly reduced footprint and complexity. In particular, for RF transversal functions, the number of wavelengths dictates the available channel number of RF time delays, and thus with microcombs, the performance of RF beamforming systems and filters can be greatly enhanced in terms of the angular resolution and quality factor, respectively. In addition, for photonic RF channelizers, with a given bandwidth for each wavelength channel, the total operation bandwidth (i.e., the maximum bandwidth of the input RF signal that can be processed) will depend on the number of wavelengths, and thus can be greatly enhanced with microcombs. Based on these advantages, a wide range of RF applications have been demonstrated, such as optical true time delays [23-25], transversal filters [25-27], signal processors [28, 29], channelizers [30, 31] and others [32-34]. Here, we review the recent advances of RF signal processing functions made possible through the use of microcombs, highlighting their potential and future possibilities.

## II. MICROCOMB GENERATION AND SPECTRAL SHAPING

The generation of microcombs is a complex process that generally relies on high nonlinear optical parameters, low linear and nonlinear loss as well as engineered anomalous dispersion. Diverse platforms have been developed for microcomb generation such as silica [10], magnesium fluoride [11], silicon nitride [12], and doped silica glass [13]. The MRRs used in the experiments reviewed here were fabricated using CMOS compatible fabrication processes, with Q factors of over 1.2 million and radii of ~592 μm and ~135 μm, corresponding to an FSR of ~0.4 nm (~49 GHz) and ~1.6 nm (~200 GHz), respectively [14, 27]. The platform used here is trade-named "Hydex" and has similar optical properties to silicon oxynitride. It combines extremely low nonlinear absorption, leading to an extremely high nonlinear figure of merit, particularly when compared to other nonlinear materials such as chalcogenide glasses [35-37], as well as a moderate nonlinearity.

To generate microcombs, the CW pump power is typically amplified and then the wavelength is swept from blue to red (Fig. 1(a)). When the detuning between the pump wavelength and MRR's cold resonance wavelength becomes small enough such that the intracavity power reaches a threshold, a modulation instability driven oscillation is initiated [17]. As the detuning is changed further, single-FSR spaced microcombs are generated. Figure 1(b) shows the observed distinctive 'fingerprint' optical spectrum achieved with a 49GHz FSR microcomb, which arises from spectral interference between tightly packed solitons in the cavity — so called "soliton crystals" [18].

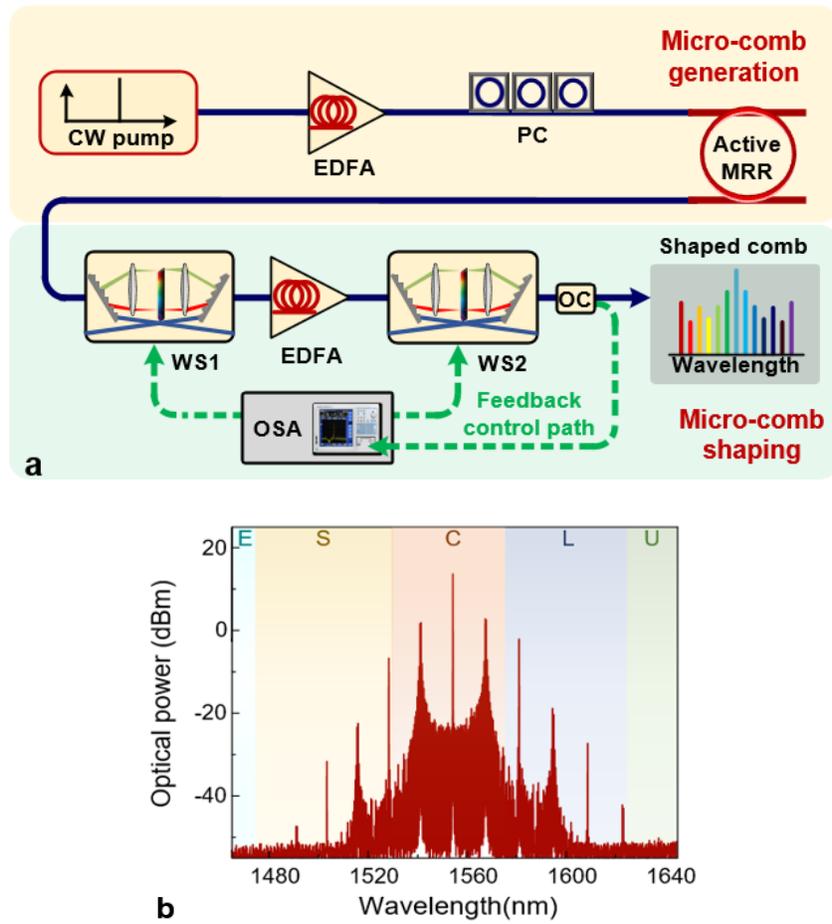

Fig. 1. (a) Schematic diagram of the microcomb generation and shaping system. EDFA: erbium-doped fiber amplifier. PC: polarization controller. MRR: micro-ring resonator. WS: Waveshaper. OC: optical coupler. OSA: optical spectrum analyzer. (b) Optical spectrum of the generated soliton crystal micro-combs with 200 nm span.

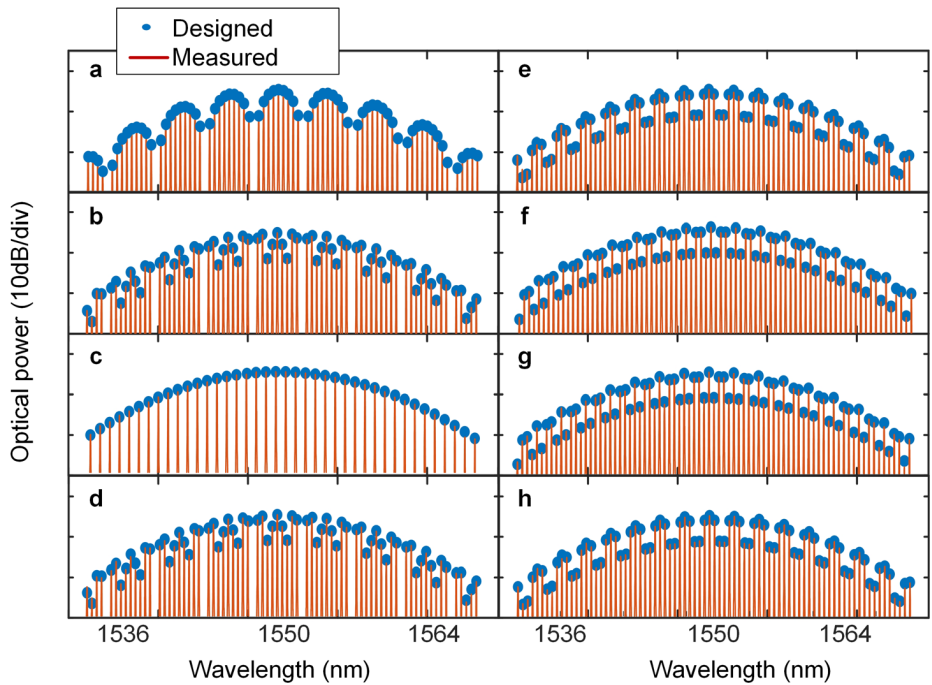

Fig. 2. Optical spectra of the shaped micro-comb corresponding to the tap weights of centre frequency tunable Gaussian apodised sinc RF filters [27].

For RF applications, the microcombs are then spectrally shaped via a two-stage optical spectral shaper (Finisar, Waveshaper 4000S) in order to enable a large dynamic range of loss control and higher shaping accuracy. The micro-comb is first pre-shaped to reduce the power difference between the comb lines to less than 15 dB, and then amplified and accurately shaped by a subsequent Waveshaper according to the designed tap weights. For each Waveshaper, a feedback control path is typically adopted to increase the accuracy of the comb shaping, where the power in the comb lines is detected by an optical spectrum analyzer and compared with the ideal tap weights in order to generate error signals for calibration. Figure 2 shows examples of shaped 49GHz comb spectra, matching well with the designed weights. The tap weights are for transversal Gaussian apodised tunable RF filters [27].

## III. RF TRANSVERSAL SIGNAL PROCESSORS

The transversal structure, similar to the concept of finite impulse response digital filters, is a fundamental tool for photonic RF signal processing. With the proper design of the tap weights, any transfer function can be arbitrarily realized for different signal processing functions such as bandpass filters, differentiators and Hilbert transformers [23-34].

As Fig. 3 shows, the core of the microcomb-based transversal structure mainly contains two operations: broadcast and then delay. The input RF signal is first broadcast onto equally spaced wavelength channels via optoelectronic modulation to generate multiple replicas of the RF input. The replicas are then transmitted through a dispersive medium to acquire wavelength-sensitive delays. Second-order dispersion, which yields a linear relationship between the wavelength and the delay, is generally employed to progressively separate the replicas. The delayed signals can be either individually converted into electronic signals or summed upon photodetection. The former can be implemented through wavelength demultiplexers and photodetector arrays to perform the function of RF true time delays [23, 24], while the latter can serve as transversal filters for RF signal processing [25-29].

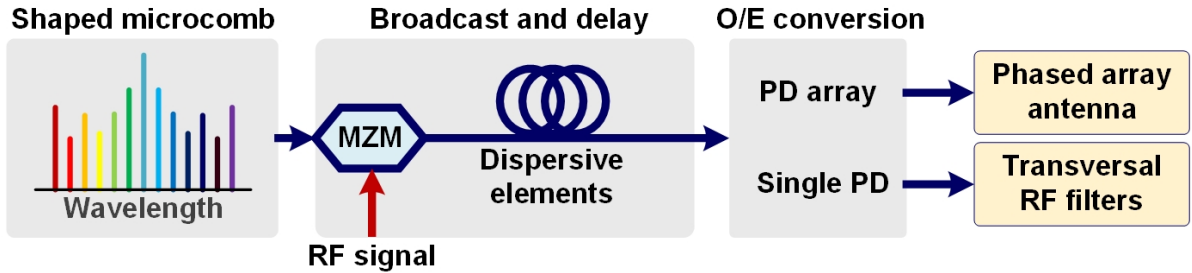

Fig. 3. Schematic diagram of the microcomb-based transversal structure. MZM: Mach-Zehnder modulator.

Photonic RF true time delays are widely employed for beamforming in phased array antennas, where the number of radiating elements determines the beamwidth [23]. Thus, in order to achieve a small beamwidth with a high angular resolution, a large number of wavelength channels is required. On-chip microcombs have the potential to provide a large number of wavelengths, while offering a small footprint and low cost, and so are promising candidates for integrated beamforming systems. Figure 4 shows recent advances made using microcombs. As the number of wavelengths increases from 21 to 81 by using a 49GHz FSR MRR, an enhancement in resolution of over 4 times is achieved, together with squint-free [21] and reconfigurable beam steering angles. Meanwhile, the large number of wavelengths offered by the microcombs also provides significant advances in the performance of transversal signal processors [28, 29]. The progressively delayed RF replicas are termed "taps", and the number of taps directly determines the resolution, or the $Q_{RF}$ factor, of the system. Figure 5 (b) shows the linear relationship between the $Q_{RF}$ factor of the transversal filter (using sinc filters as benchmarks) and number of taps.

On the other hand, the tap coefficients, scaled by the power distribution of the comb lines in the comb shaping process, determine the transfer function of the system. By properly setting the tap coefficients, a reconfigurable transversal filter with diverse transfer functions can be achieved [5], including Hilbert transformers [29], differentiators [28], and bandpass filters [25-27], as shown in Fig. 5. It can be seen that all the configurations exhibit a response that agrees well with numerical simulations, verifying the attractiveness of microcomb-based transversal approaches.

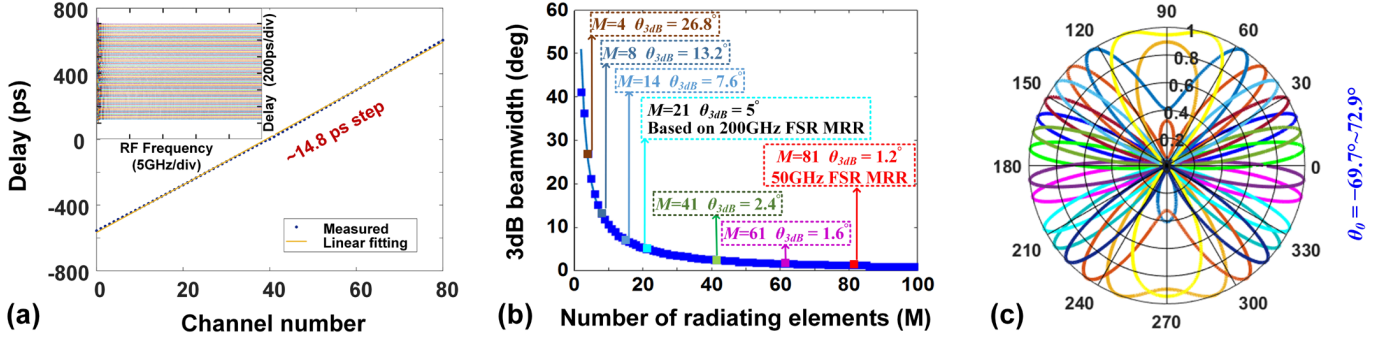

Fig. 4. (a) Measured time delay of the 81-channel RF true time delays, the inset shows flat delays over a wide RF frequency range. (b) Relationship between number of radiating elements ($M$) and the 3dB beamwidth ($\theta_{3dB}$). (c) Calculated array factors of the phased array antenna based on the 49GHz-FSR Kerr comb, with beam steering angles ranging from −69.7° to 72.9° [23].

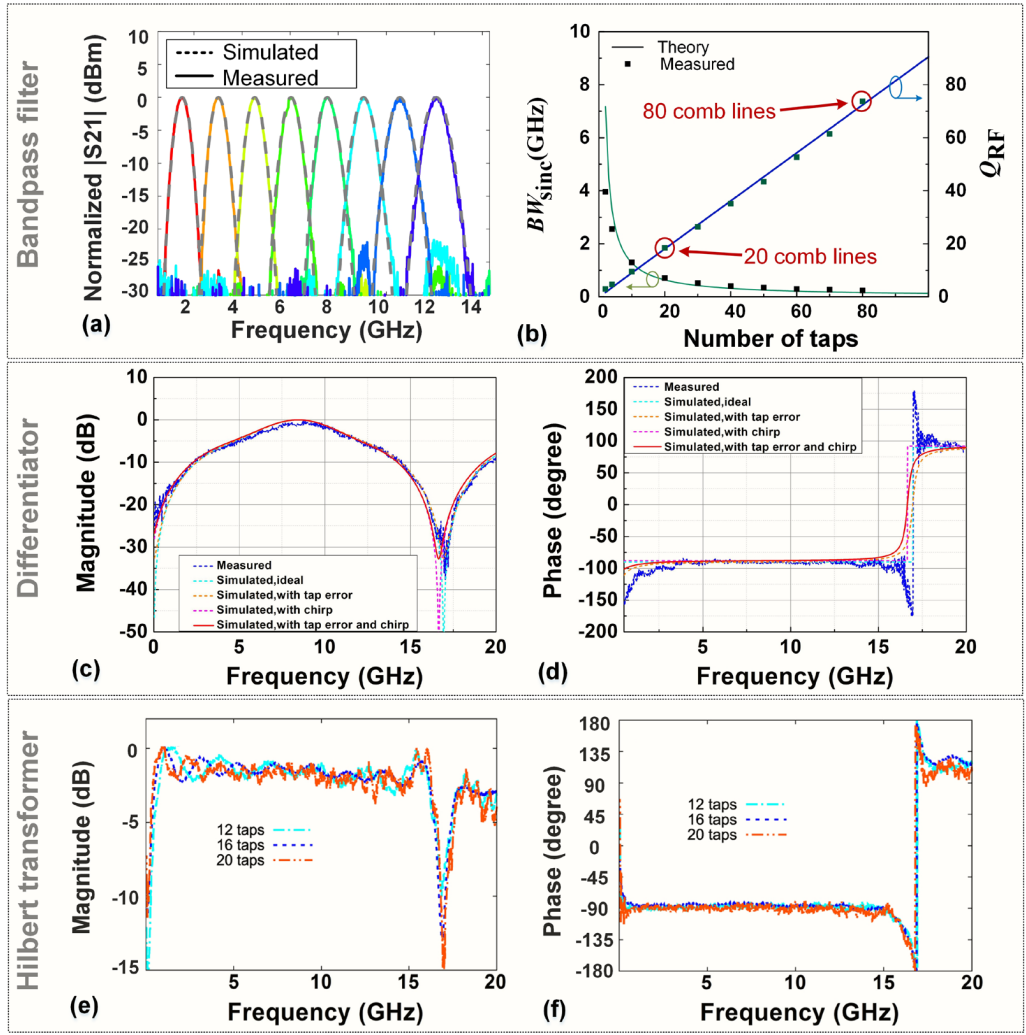

Fig. 5. Measured RF transmission spectra of (a) the bandpass filter with a tunable centre frequency, (b) the relationship between the number of taps and the $Q_{RF}$ factor, (c, d) differentiator [27,28] and (e, f) Hilbert transformer [29].

## IV. RF CHANNELIZERS

Photonic RF channelizers are another important microcomb application [30, 31]. The core concept of channelization is to slice the input broadband RF spectrum into multiple segments, such that each segment features a bandwidth within the capability of digital electronics, enabling powerful digital-domain tools to be used for processing wideband analog RF signals.

Photonic implementation of RF channelizers typically employs frequency combs to multicast the input RF spectrum onto all wavelength channels (Fig. 6), then slicing the spectrum with a periodic optical filter (such as an MRR [30]) that has an FSR slightly different from that of the comb source [31]. Finally, the wavelength channels are separated with demultiplexers and converted into the electrical domain for post digital processing.

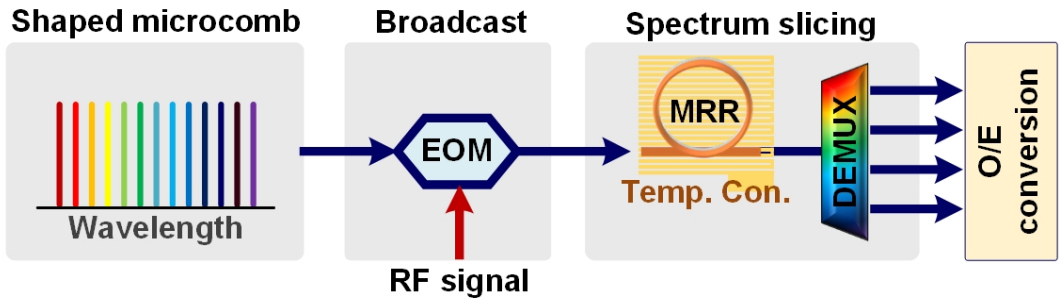

Fig. 6. Schematic diagram of the microcomb-based channelizer. EOM: Electro-optical modulator. Temp. Con.: temperature controller. DEMUX: optical wavelength demultiplexer.

A key parameter for channelizers is the overall operation bandwidth, which scales linearly with the number of wavelength channels. For example, assuming that the sliced spectra have a 2 GHz bandwidth (equal to typical electronic analog-to-digital converters) for each wavelength, then with a microcomb source that can generate 80 comb lines, an ultra-large 160 GHz operation bandwidth can potentially be reached for the channelizer—in contrast to a 10 GHz bandwidth obtained with 5 wavelengths that is the generally achievable number for discrete laser arrays.

We recently demonstrated a high-resolution RF filter using the microcomb-based RF channelization technique [31], in which the channelized RF spectral segments were manipulated in power by controlling the wavelength channel weights. Figure 7 shows the reconfigurable notch filtering shapes with binary weights. The RF transmission spectra agree well with the shaped comb spectra, illustrating the power of the channelization technique and the reconfigurability and resolution of the RF filter.

In terms of the long-term goal of full integration of these approaches to RF signal processing based on micro-combs, two key challenges are the waveshaper and the delay lines. We note that simplified versions of the waveshapers used here have been reported in integrated form [38]. Further, recent breakthroughs [39] in ultra-low loss SiN raise the prospect of fully integrated delay lines, potentially 100's of meters in length, which would be sufficient for these applications.

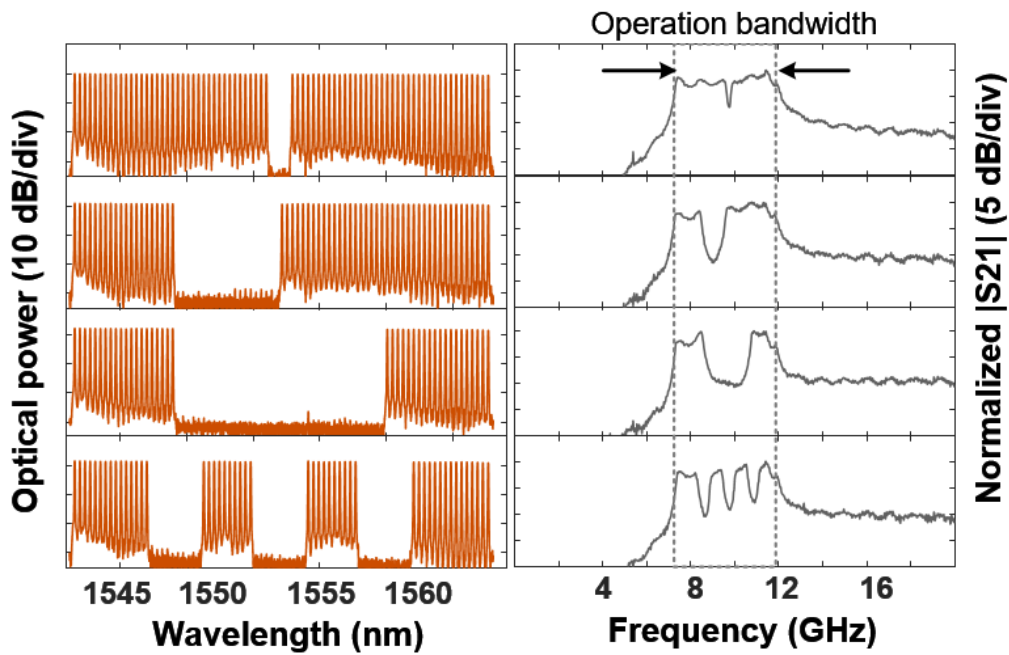

Fig. 7. Measured RF transmission spectra (right column) of the RF filter featuring binary coded (on or off) channel weights and corresponding optical spectra (left column) of the shaped micro-comb [31].

## V. Conclusion

Microcombs have shown their powerful capabilities for RF signal processing due to their large number of comb lines and compact footprint. We believe that they will bring further benefits to RF photonics, particularly in two aspects. First, the coherent nature of the soliton states [17] will enable more advanced RF functions such as wideband frequency conversion and clock generation. Secondly, globally established CMOS fabrication platforms, which can perform hybrid integration of the microcomb source and III-V devices, will potentially enable monolithic integration of the entire RF system. This will greatly enhance the performance of photonic RF systems in terms of footprint, reliability, cost and power consumption for applications such as 5G and satellite communications.